# Do We Perceive Reality?

**John Klasios**
jklasios@uoguelph.ca

Department of Philosophy, University of Guelph

## Abstract

The cognitive scientist Donald Hoffman argues that we don't perceive reality: spacetime, objects, colors, sounds, tastes, and so forth, are all merely an interface that we evolved to track evolutionary fitness rather than to perceive truths about external reality. In this paper, I expound on his argument, then I extend it, primarily, by looking at key ideas in physics that are quite germane to it. Among the topics in physics that I discuss are black holes, the holographic principle, string theory, duality, quantum gravity, and special relativity. I discuss these ideas from physics with an eye to their relevance for Hoffman's view.

## Introduction

The cognitive scientist Donald Hoffman (2019) argues that we don't perceive reality: spacetime, objects, colors, sounds, tastes, and so forth, are all merely an interface that we evolved to track evolutionary fitness rather than to perceive truths about external reality. (I will use the terms external reality and reality interchangeably; the term reality will be a shorthand for everything outside the mind of the perceiving agent.) In this paper, I expound on his argument, then I extend it, primarily by looking at key ideas in physics that are quite germane to it. Among the topics in physics that I discuss are black holes, the holographic principle, string theory, duality, quantum gravity, and special relativity. I discuss these ideas from physics with an eye to their relevance for Hoffman's view.

## Does science tell us the truth about reality?

Hoffman's theory cuts to the core of the debate on scientific realism.[1] In the philosophy of science, scientific realism, to a first-pass approximation, says that our scientific theories accurately represent reality (that is to say, the external world). Anti-realism, on the other hand, says that our scientific theories do not accurately represent reality. A well-known argument against scientific realism, most famously articulated by Thomas Kuhn (1970), comes from the history of science. Kuhn argued that we should at the very least be skeptical about the truth or accuracy of all our currently accepted scientific theories simply because the accepted scientific theories of the past have eventually been shown to be false and replaced with successors. According to this argument, we should, given enough time, expect all of our currently accepted scientific theories to suffer the same fate.

Another form of anti-realism goes by the name instrumentalism, which says that we should only be concerned with whether our scientific theories (or theories and explanations in general) make accurate predictions. Whether our theories are in some sense accurately representing reality or not, the only thing we can know about and indeed should care about, according to instrumentalism, is whether theories make accurate predictions. An influential version of instrumentalism, developed by the philosopher Bas van Fraassen (1980), is called "constructive empiricism". Constructive empiricism says, essentially, that we should only believe in the existence of entities that can be observed by the naked eye. Any entities posited by a scientific theory which are not observable by the naked eye are entities we should be agnostic about: we can't know whether they actually exist or not. According to constructive empiricism, the entities that are unobservable to the naked eye, when they are part of scientific theories, are useful for trying to make accurate predictions, and that's it. Constructive empiricism says we can wonder about whether the unobservable entities in a theory actually exist in external reality, but this wondering is just idle speculation that cannot be decided one way or another.

[1] Some other treatments of the debate on scientific realism can be found in Chakravartty (2017), Rosenberg (2011), and Godfrey-Smith (2009). Balashov and Rosenberg (2001) is a compendium of landmark papers in the philosophy of science.

Realists have responded to all of these objections. In response to constructive empiricism, they have argued that the boundary between the observable and the unobservable is artificial and unstable: science is continually shifting the line that demarcates what's observable from what is unobservable. For instance, at some point in the past, neither the surface of Mars nor distant galaxies was unobservable. But technology has made these observable. Similarly, realists have argued that being able to detect things and being able to observe them are more closely associated than they might otherwise seem. After all, the vision that we have with the naked eye is a kind of detection too. Accordingly, it might be more sensible to think that while we should be more cautious about inferring that we're accurately representing reality when we detect it with means other than vision, nonetheless standard vision isn't some kind of insuperable barrier that prevents us from accurately representing reality beyond that point.

Realists have also responded to the argument that we should not believe our best current theories because the best current theories of the past, as well, have been shown to be false and ultimately replaced. Realists have argued that realism is probably true because, if it weren't, the only way that scientific theories could predict in the accurate way that they do would be by some miraculous coincidence; because how else could they accurately predict if they were not in some sense accurately representing reality (Putnam, 1975)? Anti-realists have responded to this rejoinder by saying that there is no need whatsoever for the entities of a theory to correspond to real entities in reality for the theory to make accurate predictions. Sadi Carnot, for instance, gave an explanation of heat in terms of a fluid rather than in terms of the motions of molecules; yet his explanation predicted heat well not because heat is actually a fluid, but because his explanation captured the right patterns that underlie the behavior of heat (Godfrey-Smith, 2009). So according to anti-realism, there is no guarantee that the entities posited by a theory exist in reality just because the theory makes accurate predictions.[2]

Realists, in turn, have countered these kinds of objections with a view called "structural realism" (Worrall, 1989; Ladyman et al., 2007). Accordingly, when any past theory in the history of science—especially in physics—has been shown to be "false"—or rather, strictly speaking false—and replaced by a successor, the successor theory has nonetheless retained an abstract, structural core from the previous theory. In physics, the core that is retained is mathematical. So, the structural realist would thereby say that though it might superficially seem that theories are shown to be false by the theories that succeed them, even the theories that are ultimately replaced nonetheless contain some structural core that accurately represents reality. For example, some degree of mathematical structure in one theory in physics is retained by a theory that replaces it; and, should that replacement theory in turn be replaced by a theory later on, the idea is that some degree of its mathematical structure will also be retained, and such that mathematical structure can be retained over time in a cumulative way. Thus, there can be a kind of scientific progress, such that science comes to represent reality in an increasingly accurate way. So structural realists argue that we can "get closer to the truth" over time. More concretely, and to use an example, this helps to explain why Einstein, despite overturning Newton's physics, nonetheless thought he was building upon Newton's work and retaining some amount of its content. And more practically, this is also why the Apollo missions to the moon were able to utilize Newton's equations rather than those of relativistic physics, even though the latter theory is, strictly speaking, more accurate and the former theory's equations are, strictly speaking, false. Despite the fact that Newton's theory is technically false, his equations nonetheless accurately capture aspects of reality—and in the case of gravity, they accurately describe its behavior within the domain of ordinary scales and velocities.

[2] Philosophers of science often refer to this idea as the "under-determination of theory by the evidence", or more simply as the "under-determination thesis". Essentially, the contention is that any given observation, or set of observations, is consistent with more than one explanation (more than one theory), perhaps uncountably many. A discussion of these issues in the context of the history of science can be found in (DeWitt, 2018).

There are also two versions of structural realism that will be of interest to us in light of Hoffman's theory, as will be shown. "Ontic structural realism" says that reality is, fundamentally, made up of structure—such as logical or mathematical relations—and nothing more. On the other hand, philosophers of science debate whether it even makes sense in principle to say that pure structure, and nothing more, is what reality fundamentally is. For instance, it's been argued that for there to be relations, there must be "things"—entities of some kind—to stand in those relations in the first place (e.g., Cao, 2003). The basic idea here is that there must be some concrete entities of some kind for there to be anything real—for example, for a planet or an organism to exist, there must be something tangible like quarks or molecules that make up the planet or organism. Without those tangible entities, so the criticism goes, there simply cannot be any relations—the relata come first, and the relations supervene on them. Later on, when we speculate on what reality might be like in light of Hoffman's theory, we'll consider a version of ontic structural realism, namely the view that reality is fundamentally mathematical (Tegmark, 2014).

"Epistemic structural realism", on the other hand, says that there may or may not be entities that stand in relation to one another in some way, and which comprise reality, but the only kind of knowledge we can have, in principle, is knowledge of structure. For instance, whether or not there are electrons and hadrons, we can't know; but we can know about the mathematical structure that underlies the phenomena that we refer to in our talk about electrons and hadrons and their behaviors. Hoffman's theory, if true, would strongly challenge realism of any kind. If it were true, we might very well be unable to say whether reality is or is not made up of pure structure of some kind (i.e., logical or mathematical); likewise, if Hoffman's theory is true, we might not be able to have even structural knowledge of reality. Because if Hoffman's theory is correct, we inhabit, in the first place, a kind of video game that evolved to serve as our perception of reality, whatever that reality is. In such a predicament, whether we can then probe "beyond" the metaphorical video game in order to understand external reality becomes an open question.

Before proceeding, it will be useful to make a brief clarification on the notion that science can accurately represent reality. The "correspondence theory of truth" is a traditional idea in philosophy that says that a theory or explanation can depict reality in the way that a picture can depict a scene. But philosophy of science has generally moved past this idea and now embraces a different kind of view with regard to realism. In its place is the view that scientific theories can represent reality in any of a number of different formats. For example, a theory can represent reality with a model of some kind, or using mathematical equations, or with language, and so on. So, strictly speaking, scientific theories do not necessarily need to "picture" reality in some intuitive way and can instead come in the form of various kinds of abstract representational formats (Godfrey-Smith, 2009).

## Fitness beats truth

Let's start with the main building blocks of Hoffman's argument. Evolutionary game-theory is a way of formally modeling the hypothetical interactions between organisms. For example, a model in game theory might look at what happens when a population of hawks and doves interact: the model can stipulate what percentage of the population is made up of hawks and doves, how many interactions there are between hawks and hawks, between doves and doves, and between hawks and doves, and how many points a hawk or dove gains or loses when it interacts with another hawk or dove. Doves don't fight when they encounter a hawk, so they lose points; doves make peace when they encounter other doves, so each gets some points; hawks fight when they encounter another hawk, so each loses points half the time and gets points the other half of the time (and sometimes the hawk that wins a fight against another hawk still gains less points than when they encounter doves, since even a losing hawk can inflict damage on the other). Using the model, we can tally up how many points a hawk and dove end up with, overall, after all interactions. For instance, some models show that doves end up better off than hawks if the percentage of hawks in the population exceeds a certain number; but those same models show that hawks fare better than doves if the percentage of hawks in the population is lower than a certain number. In these models, the percentages of hawks and doves stabilize at certain proportions—for instance, 20 percent hawks and 80 percent doves. Evolutionary game-theory can thus be used to model how different traits and strategies fare in terms of reproductive fitness.

Using the logic of evolutionary game-theory, Chetan Prakash and Hoffman proved a theorem that they called the *Fitness-Beats-Truth Theorem* (henceforth FBT Theorem) (Hoffman & Prakash, 2014; Prakash et al., 2021; Hoffman et al., 2015). In the theorem, two sensory strategies are possible, one that sees the *truth* about the external world as best as possible, and the other that sees only evolutionary *fitness* (essentially, whatever gives an advantage in reproduction). Formally, according to the FBT Theorem, "*fitness* drives *truth* to extinction with probability at least (*N*-3)/(*N*-1)" (Hoffman, 2019, p. 82, emphasis in original). Hoffman (2019) explains the theorem this way:

> Consider an eye with ten photoreceptors, each having two states. The FBT Theorem says the chance that this eye sees reality is at most two in a thousand. For twenty photoreceptors, the chance is two in a million; for forty photoreceptors, one in ten billion; for eighty, one in a hundred sextillion. The human eye has one hundred and thirty million photoreceptors. The chance is effectively zero. (p. 82)

To further illustrate the FBT Theorem, consider a simulated world with two types of beings, Blobs and Philosophers. Both subsist on rice, and rice is distributed in the world in small, medium, and large concentrations. Medium concentrations deliver fitness points, but large and small concentrations deliver no fitness. But whereas Philosophers perceive all the truth about the world, Blobs are tuned to perceive what's relevant to fitness. Thus, Philosophers perceive blue where rice is in small concentrations, green where rice is in medium concentrations, and red where rice is in large concentrations. Blobs, by contrast, perceive blue where rice is in either small or large concentrations, and red where rice is in medium concentrations. So, Philosophers track rice distribution in a more accurate way than Blobs: blue when less rice is there, green when it's a moderate amount, and red when more of it is there. Blobs don't track all the truth, but they lock onto fitness directly: the fitness-boosting rice—the rice in medium concentrations—is salient to them, but not in the same way to Philosophers. So, Blobs survive and Philosophers go extinct. Fitness-finding survives, truth-tracking goes extinct.

This simulation is obviously very simple. But Hoffman points out that even when we start to increase the complexity of the situation, and even when we include other evolutionary forces such as population size, mutations (including neutral mutations), and genetic drift (where chance events can affect the makeup of the gene pool), fitness still appears to have the edge over truth. Justin Mark (2013; see also Hoffman et al., 2013; Mark et al., 2010) ran simulations in an artificial world to further test the FBT Theorem. In these simulations, populations made up of a finite number of agents had their perception and behavioral strategies co-evolve over many "generations". Initially, perception and behavioral strategies varied between the agents. After a generation, the ones that ate the most resources had their genes copied into the next generation, and some of their genes were randomly mutated in order to provide new variation. This process was iterated for hundreds of generations. In the last few generations, the surviving agents only perceived the fitness of resources, but not the real quantities of—the truth about—the resources.

## The interface theory of perception

Another one of the main building blocks of Hoffman's argument is the Interface Theory

of Perception (henceforth ITP).[3] Because Hoffman thinks that agents evolve to see fitness rather than truth, he argues that the format of perception that they end up evolving is like the operating system of a computer. In our case, we have a desktop populated by icons. The desktop is analogous to the space and time that we perceive, and the icons are analogous to the objects we perceive, along with the properties of those objects, such as shape, color, texture, and smell. The operating system that we evolved is a useful way of tracking fitness: it organizes reality, whatever reality might be, using a user-friendly interface. When we perceive a chair, for instance, it is like seeing an icon on the desktop of an operating system. The chair no more resembles external reality than the icon resembles the transistors, voltages, programming language, and such, that realize the icon; the icon is merely a useful but radically different way of representing that more complicated reality.[4] Entities that are not directly perceivable but that we can in principle perceive, such as molecules and electrons, are also mere icons; they don't literally exist in external reality. Galileo and Locke made a distinction between "primary" and "secondary" qualities: primary qualities such as solidity, mass, and motion are those that objectively exist in external reality, whereas secondary qualities such as sound, color, and taste only exist in the mind of the perceiver. But Hoffman's view gets rid of the whole kit and kaboodle: reality not only lacks properties like sound, color, and taste, it also lacks objects, and even space and time.

Hoffman's view has a clear parallel with the Kantian distinction between the "things-in-themselves"—the noumena—and how we perceive reality—the phenomena. In our usual way of thinking about visual perception, species evolve to see only certain parts of external reality; we don't, for example, directly perceive molecules or quarks (and other sub-atomic particles). Seeing that much detail about external reality would be unnecessary and potentially distracting, and would likely not enhance reproductive fitness. This is also what the game-theory-based simulation with Philosophers and Blobs indicates. As Hoffman sees it, when no agent perceives an apple or a quark, the apple or quark *as such* does not exist. Whatever corresponds to the apple or quark in external reality continues to exist, but the perceptual icon—namely the apple or quark—vanishes until it is recreated in someone's operating system through perception. It is in this specific sense that Hoffman's view also echoes Berkeley's famous argument that "to be is to be perceived". But again, we should reiterate that whatever corresponds to the apple or quark in reality continues to exist, even when not perceived; we just don't know what their real nature is.

According to the ITP, two key advantages of perceiving reality in terms of space, time, and objects are data compression and error-correction. The ITP argues that perceiving in terms of spacetime and objects is a form of data compression: a way of compactly and frugally representing the fitness-relevant properties of external reality. The ITP is thus similar in a sense to Gibson's (1986)

[3] Our summary of Hoffman's (2019) view will be concise. The interested reader is encouraged to read his book for more details and fuller context.

[4] Hoffman's view echoes pragmatist views such as Rorty's (2009).

“affordances”, in that what we perceive most directly are what might give us fitness points: food, mates, ways to act, etc. Similarly, it parallels Jakob von Uexküll’s concept of an umwelt: a kind of perceptual world that a species evolves, which makes certain things and phenomena in its environment salient because of their importance to the organisms in that species (Agamben, 2004). In the game of evolution, all that ultimately matters is surviving long enough to produce viable kids. As long as agents have a fitness edge over the competition, they, and their genes, will make it to the next round; so perceptual representations need only satisfice better than the competition’s perceptual representations. The ITP also argues that perceiving in the format of spacetime and objects is a form of error-correction: it injects redundancy such that we represent reality in a way that is robust to noise, to error. To illustrate error-correction, imagine meandering through a forest shrouded in dense fog at night. What you see in your visual field are blurred edges, sub-optimal lighting, and lots of ambiguous shapes. The information that you perceive, in other words, is noisy and fairly incomplete. Still, instead of seeing a “blooming, buzzing confusion”, in the words of William James, you see stable objects in your visual field. The same tree that you see from a distance is still recognized by you as the same tree as you walk by it, and you likewise recognize some leaves by your feet as the same leaves even if your vantagepoint and their luminosity change. This way of representing what you see is a form of error-correction. The noisy and incomplete information arriving at your senses is interpreted as if there are stable objects in your environment. Error-correction thus buttresses perception by making it more reliable despite distortive and changing input from the environment.

Before we move on to physics, we should consider one important objection, namely whether the FBT Theorem and ITP are self-refuting. Does Hoffman’s view hoist itself by its own petard? Hoffman responds by correctly pointing out that the very idea of evolution via natural selection is abstract. Richard Dawkins (1983) has pointed out that the abstraction of natural selection lends itself to “Universal Darwinism”, the idea that natural selection can apply to a wide variety of entities. This is because evolution by natural selection is substrate neutral: it can apply to any set of entities that can realize heritable variation and filtration. For our purposes here, it’s very important to underscore that the idea of natural selection is wholly abstract. In principle, then, it may apply to entities that exist without being embedded in space and time: for example, it may apply to entities that are pure mathematical constructions, which will be an idea that we’ll consider later. So, if the FBT Theorem and ITP are both right, it can still be the case that Universal Darwinism applies to external reality, even a reality sans space and time. More generally, the stance taken by Hoffman is that logic and mathematics can still function as a gateway to helping us understand what external reality might actually be like—and so we might be able to pierce the veil of perception and grasp the Kantian noumena. No doubt, the idea that logic and mathematics are left largely or entirely unscathed by the FBT Theorem and ITP deserves further scrutiny.

## Black holes and holography

As I mentioned at the outset, physics can help buttress Hoffman's theory. A good place to start is with black holes. Early in the 20th-century, the physicist and astronomer Karl Schwarzschild developed the first precise, fully worked-out solutions to Einstein's then-nascent theory of gravity, general relativity. Schwarzschild showed precisely how spacetime would be warped and twisted around relatively large bodies such as a planet or a star. But he also showed what would happen in a region of space whose fabric had been so warped and twisted that not even light could find its way out from such a distortion of space—which would occur if a sufficiently large amount of mass had been concentrated in that region. In the late 18th-century, Pierre-Simon de Laplace and John Michell, reflecting on Newton's physics, wondered whether there could be stars so massive that even light couldn't escape their gravitational pull, and thus would appear black to observers on the outside (Susskind, 2008). This hypothesis would later be borne out in a precise, mathematical way by Schwarzschild, whose solution described just such a star. According to Schwarzschild's solution, if enough mass was packed into a region of space, nothing, not even light, would be able to escape its pull. Schwarzschild provided a formula that could ascertain where the exact boundary around such an object would be: namely, the surface beyond which incoming light would not be able to escape the gravitational pull, known as *Schwarzschild's radius*. For example, picture an object so massive that it distorts the fabric of space in the way Schwarzschild calculated. Now picture orbiting the object at a safe distance and throwing in a flashlight towards it and past the boundary surface found at the Schwarzschild radius. The gravitational well created by the massive object would be so strong that light from the flashlight could not escape and be seen by you on the other side of the boundary surface. The boundary surface found at the Schwarzschild radius is more commonly referred to as the event horizon.

**Figure 1**

The Schwarzschild Radius of a Black Hole

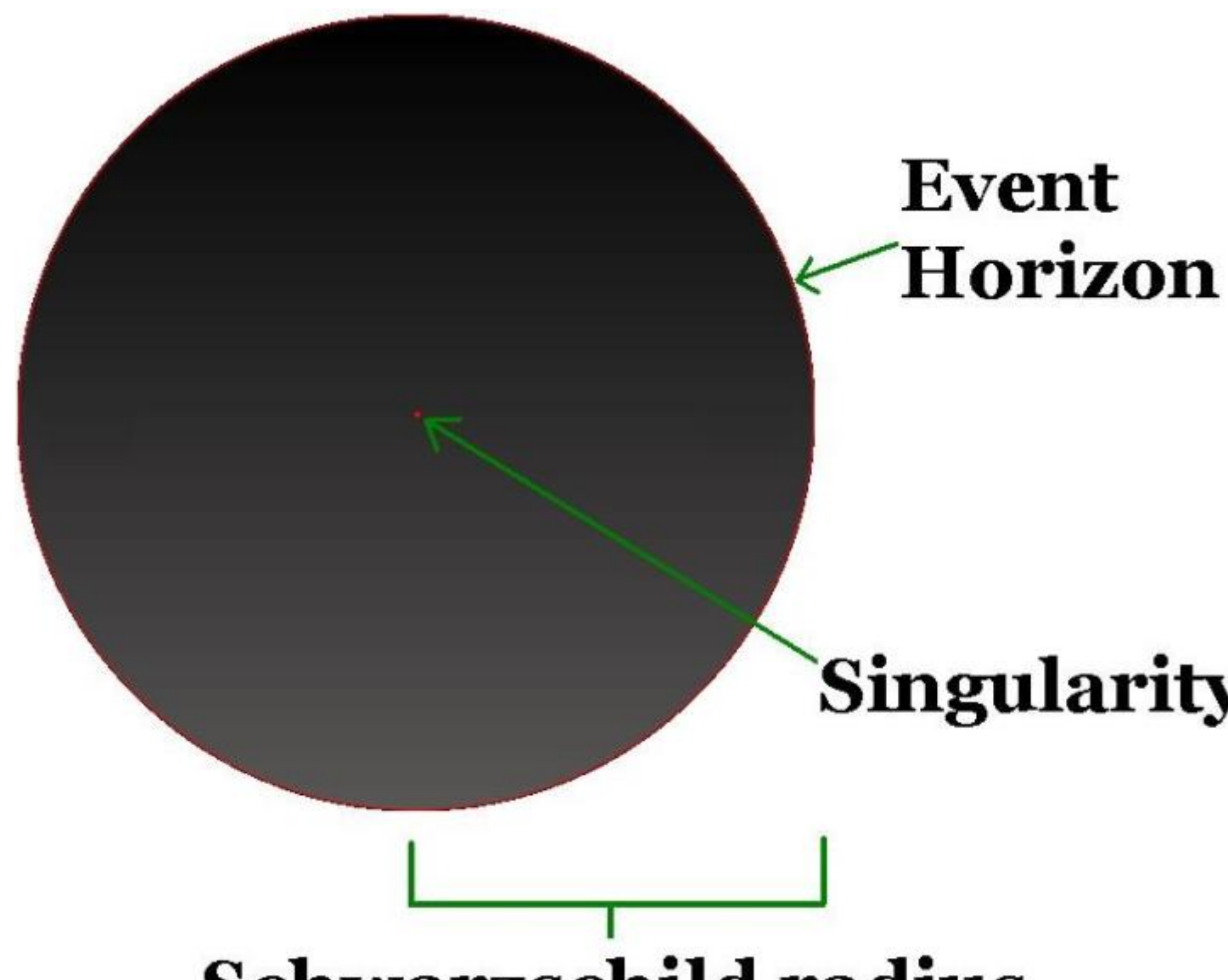

*Note*. The Schwarzschild radius of a black hole is the distance between the hole's singularity at the center and the event horizon at its surface. From *Black hole details* by T. Quark (2021), Wikimedia Commons ([https://commons.wikimedia.org/wiki/File:Black_hole_details.JPG](https://commons.wikimedia.org/wiki/File:Black_hole_details.JPG)). Licensed under **CC BY-SA 4.0**.

Far from being just a mere theoretical oddity, black holes have been amply confirmed by astronomers—a picture of one has even been taken (Landau, 2019). For example, massive black holes are found at the center of galaxies. Indeed, there's a massive black hole at the center of our own Milky Way galaxy, one with a mass of about 3 million times the mass of our own sun. Gravitational waves produced by colliding black holes have even been detected by very sensitive apparatuses.

Later in the 20th century, physicists started to ask what would happen to the information that entered a black hole. Would it increase the entropy—the disorder—of the hole? The theoretical physicist Jacob Bekenstein (1973) pondered this issue and used the following thought experiment to set the stage. What would happen if a container of gas, which contained a specific amount of entropy—essentially, an amount of disorder—was flung into a black hole? Prior to Bekenstein's work, the consensus among physicists was that black holes were a kind of garbage compactor that would squeeze any matter that went in into an orderly mass at the center. This belief was founded on the understanding of black holes given by general relativity, according to which all black holes can be described with three properties, namely their mass, charge, and angular momentum (Susskind, 2008). If you ascertain these properties, then you've specified all there is to know about a given hole. According to general relativity, all matter that went into a black hole would be pulled in towards the center, where spacetime is infinitely distorted at a single point, a so-called "singularity". The renowned physicist John Wheeler (who also happened to be Bekenstein's PhD advisor) characterized this putative fact as black holes having "no hair", the idea being that the surface of black holes are featureless (with each hole being completely specifiable by mass, charge, and rotation).

Bekenstein, however, argued that a black hole also has entropy, and that tossing a container of gas into the hole would indeed increase its entropy. To demonstrate this theoretically, Bekenstein asked what would happen to a black hole if only one piece of information could be added to it. In particular, he asked what would happen if the only change made to the hole was whether that piece of information was inside or not, and if it were impossible to know the specific location on the surface at which it entered the hole. The technical details of Bekenstein's argument are involved, but the basic idea is this. By taking a photon whose wavelength was precisely the size of a black hole's radius (which, as we've seen, can be calculated using Schwarzschild's formula), it can be added to the hole such that only that piece of information enters the hole. So, the photon either enters the hole or remains outside of it. According to Bekenstein's calculation, the radius of the hole would increase by $10^{-72}$ meters—an astoundingly small increase, to be sure. Another part of Bekenstein's calculation is more interesting for our purposes, however. When we ask how much the photon's mere presence in the hole will increase the area of its surface, the answer is $10^{-70}$. This is especially interesting, because this value is strangely equivalent to *one square Planck-unit*. In physics, the Planck-length (which is one side of a square Planck-unit) is often thought to be the smallest distance that space can be subdivided into.

Bekenstein's argument seems to demonstrate that the postulated fundamental unit of space, the square Planck-unit, has an uncanny connection to the amount of *information* that can be contained in a fundamental unit of space. Or, put another way, Bekenstein's argument shows that *a single unit of information and a single square Planck-unit are equivalent in some deep sense: one unit of information per Planck-unit*. Bekenstein's view was placed on a much more rigorous footing and made more precise by Stephen Hawking (1975). Hawking, who initially disbelieved Bekenstein's argument, found that black holes actually have a temperature, and thus they have entropy. If black holes have entropy, that means they can't be fully informationally specified by merely determining their mass, charge, and rotation. Hawking showed what would happen around the event horizon of a black hole if quantum-mechanical considerations were added to the mix—as indeed they must, since both quantum considerations and relativistic considerations must both be at play in black holes. For an observer who crosses the event horizon, any pair of particles that spring forth via a fluctuation in a quantum field very close to the event horizon will look to them a certain way: one will have positive energy and the other will have negative energy. But, oddly enough, a particle that had negative energy to someone watching outside the hole would have positive energy to someone inside the hole (Greene, 2011).

From the perspective of the observer inside the hole, the particle with negative energy would stream towards the center of the hole, and, because it has negative energy, would decrease the size of the hole. To the observer outside the hole, the other particle from the pair would look like *Hawking radiation*, which means that the hole would also have a temperature. Hawking thus provided a rigorous connection between the information of a black hole and the temperature of a black hole, which showed that black holes have entropy, after all. And he showed that the amount of information contained by a black hole—its total entropy—is precisely equivalent to the surface area of the black hole (its event horizon): one unit of information equals one square Planck-length. To get a visual idea of how information is contained by a black hole, picture a very tiny square tile whose size is exactly equal to a square Planck-length. Now picture the event horizon of a black hole being covered by these square tiles. Each tile represents only one unit of information—one bit—and the black hole, depending on the precise size of the event horizon, can contain no more information than what can be represented by all the tiles.

't Hooft (1993) and Susskind (1995) further developed this view of black holes. Here's a vivid illustration of part of what they found. Picture Nicole in space headed towards a black hole, with Ghazala observing her from far away. From Nicole's point of view, she safely crosses the black hole's event horizon uneventfully; in fact, because of Einstein's equivalence principle, she knows, since her trajectory is relenting to the full force of the hole's gravity, that her journey past the event horizon will feel no differently than if she were floating motionlessly in empty space far from the black hole. But what would Ghazala see from afar? We might assume that she'd see exactly what Nicole saw. But things would be much stranger. In fact, Ghazala, after recording the light coming from the hole and from Nicole, would calculate, using Hawking's insights, that Nicole erupted into an inferno and was reduced to ashes when she reached about one Planck-length's distance from the event horizon of the black hole. As such, Ghazala calculates that the information comprising Nicole, as instantiated by her particles, would get scrambled at the event horizon. (Even though the light recorded by Ghazala would have an extremely low temperature, she knows from Hawking's equations that the light emanating from the event horizon has been fighting off the pull of gravity from the black hole as it makes its way towards her. Hence, the light's temperature would've been orders of magnitude hotter when it was very close to the hole—especially at one Planck-length's distance from the event horizon, which she calculated was the spot of Nicole's fiery demise.) So, we're left with two blatantly contradictory accounts of Nicole's fate: one where she passes through the event horizon uneventfully, and one where she dies instantly and her information is scrambled. How can the paradox be resolved?

Susskind (2008) argued that, much like the so-called wave-and-particle duality of quantum mechanics, both Nicole's and Ghazala's perspectives are equally valid. But because it would be impossible for Nicole to escape the black hole after passing the event horizon in order to go back and tell Ghazala that nothing happened after she passed it, and because, by the time Ghazala had recorded the death of Nicole near the event horizon it would be impossible to travel to Nicole and give her evidence of her death, physics ensures that there's no paradox: Nicole can't escape the pull of the black hole's gravity, and, because, as per special relativity, no object or signal can travel faster than the speed of light, Nicole will have been crushed at the center of the black hole long before Ghazala can either catch up with her or send a transmission to her. Strange as it may seem, then, the information comprising Nicole somehow is entirely stored at the event horizon—the surface area—of the black hole, even though Nicole experienced nothing out of the ordinary after crossing it.

Finally, we can cap-off our discussion of black holes with the punchline. Susskind (2008) argues that what's true about black holes is true about the universe in general. To illustrate, imagine any region of space and begin placing stuff—matter and energy of any sort—into it. Without changing the size of the region, if you could manage to continually pack it with stuff, eventually the stuff contained in the region would be so massive that a black hole would form. As we've seen, that's because the most stuff—the most information—that can be contained in a black hole is equal to its surface area (its event horizon). Not its volume, that is, but rather its area. And the same logic of black holes—that their information is contained by their surface area—applies no differently to *any* region of space. So, there is a real sense, then, in which our universe is a *hologram*: a three-dimensional projection of information residing on a distant surface area (see Figure 2). A typical hologram is an information pattern stored by black-and-white squiggles on a two-dimensional surface which, when light is passed through it in the right sort of way, can produce a three-dimensional image. In passing the light through the two-dimensional surface, the encoded information can be reconstructed into a three-dimensional image. *According to the holographic principle, our three-dimensional universe is like a holographic movie encoded by information stored and processed by a kind of quantum computer residing on a distant two-dimensional surface.* And this surface surrounds our universe, either very far away from us or infinitely far away (Susskind, 2008).

**Figure 2**

A Schematic Representation of the Holographic Principle

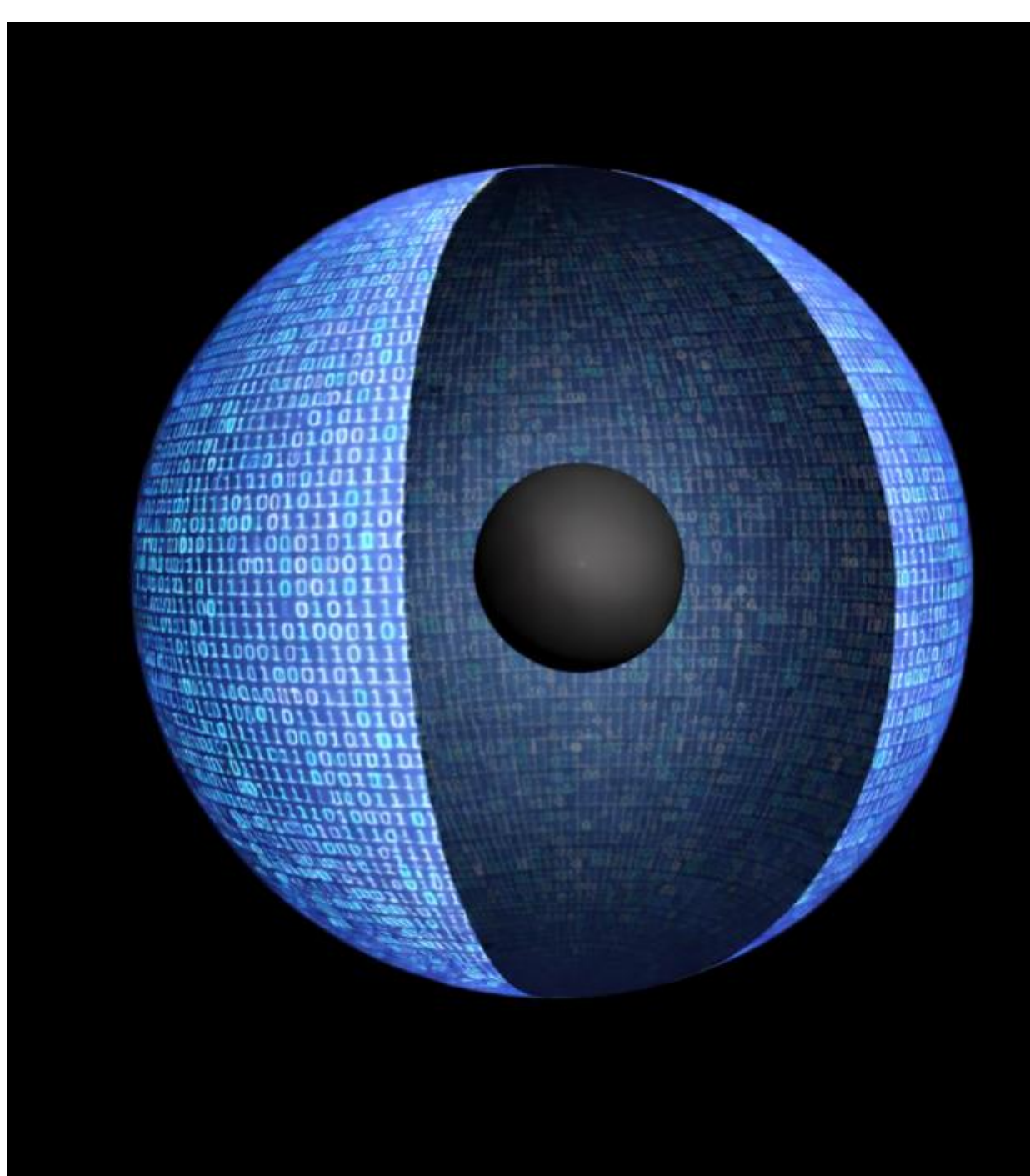

*Note*. A schematic representation of the holographic principle. All information in the 3-dimensional world resides on the 2D-surface. Objects "inside" are merely holographic projections of information on the surface. Adapted from *Deriving Emergent gravity from microscopic theories* by Yormahmad Kholov (2021), Wikimedia Commons (https://commons.wikimedia.org/wiki/File:Deriving_Emergent_gravity_from_microscopic_theories.jpg). Licensed under **CC BY-SA 4.0**.

## Holography and duality

The holographic principle emerged through the study of black holes. There we saw that the information within a region of space is found not within its volume (its interior) but rather on the surface that surrounds it. But physicists have explored ways of making the information that resides on the bounding surface tangible. A breakthrough to this end was made in the late 90s by the physicist Juan Maldacena (1999)—whose work was a watershed event that spawned countless papers that further explored the holographic principle. In order to understand the main points of Maldacena's discovery, we first need to sketch some of the basic ideas of string theory, a candidate theory for reconciling quantum mechanics with general relativity. String theory is controversial yet is the best-developed candidate theory for resolving the inconsistencies between the fundamental pillars of modern physics, namely quantum mechanics and general relativity. As it currently stands, the most developed version of string theory is known as M-theory. The main ingredients of M-theory are open strings, closed strings, extra dimensions of space, branes, and fluxes (Greene, 2011). Open strings are snippets of vibrating strings, and closed strings are loops of string (which resemble an elastic band).

M-theory also posits the existence of six extra dimensions of space, on top of the usual three of our common experience, bringing the total number of spatial dimensions to nine. The six extra dimensions are incredibly small and curled up—they are so small that they are close to the Planck-length in size ($10^{-33}$ centimeters). And far from being arbitrary posits of string theory, the six extra dimensions are what make the mathematics of the theory internally consistent and capable of potentially explaining all of the forces of the universe, including gravity. Fluxes are line-like filaments that can run through any of the holes found in the extra dimensions. Branes are another object of M-theory, and they too can come in a number of different dimensions: so-called 1branes are 1-dimensional objects that are like point particles; 2branes are 2-dimensional objects that are flat sheets; and 3branes are 3-dimensional objects that are like our ordinary 3-dimensional space. Branes can also come in higher dimensions (4, 5, 6, etc.).

To illustrate Maladecena's discovery, I will use Susskind's (2008) and Greene's (2011) exposition of it. Maldacena used what's known as an anti-de Sitter space, which is a space of any number of dimensions with negative curvature. An anti-de Sitter space has a saddle-shaped surface that extends in every direction. If a triangle was located in an anti-de Sitter space, its angles would sum to less than 180 degrees. (In a flat space with no curvature, the angles of a triangle sum to exactly 180 degrees; and the angles of a triangle sum to greater than 180 degrees in a space with positive curvature, like on a sphere.) A 2-dimensional anti-de Sitter space (henceforth referred to as AdS) would be similar in some ways to Escher's painting Circle Limit IV, where the angels and demons become increasingly small the closer they are to the outer edge of the painting (see Figure 3).

**Figure 3**

Circle Limit IV

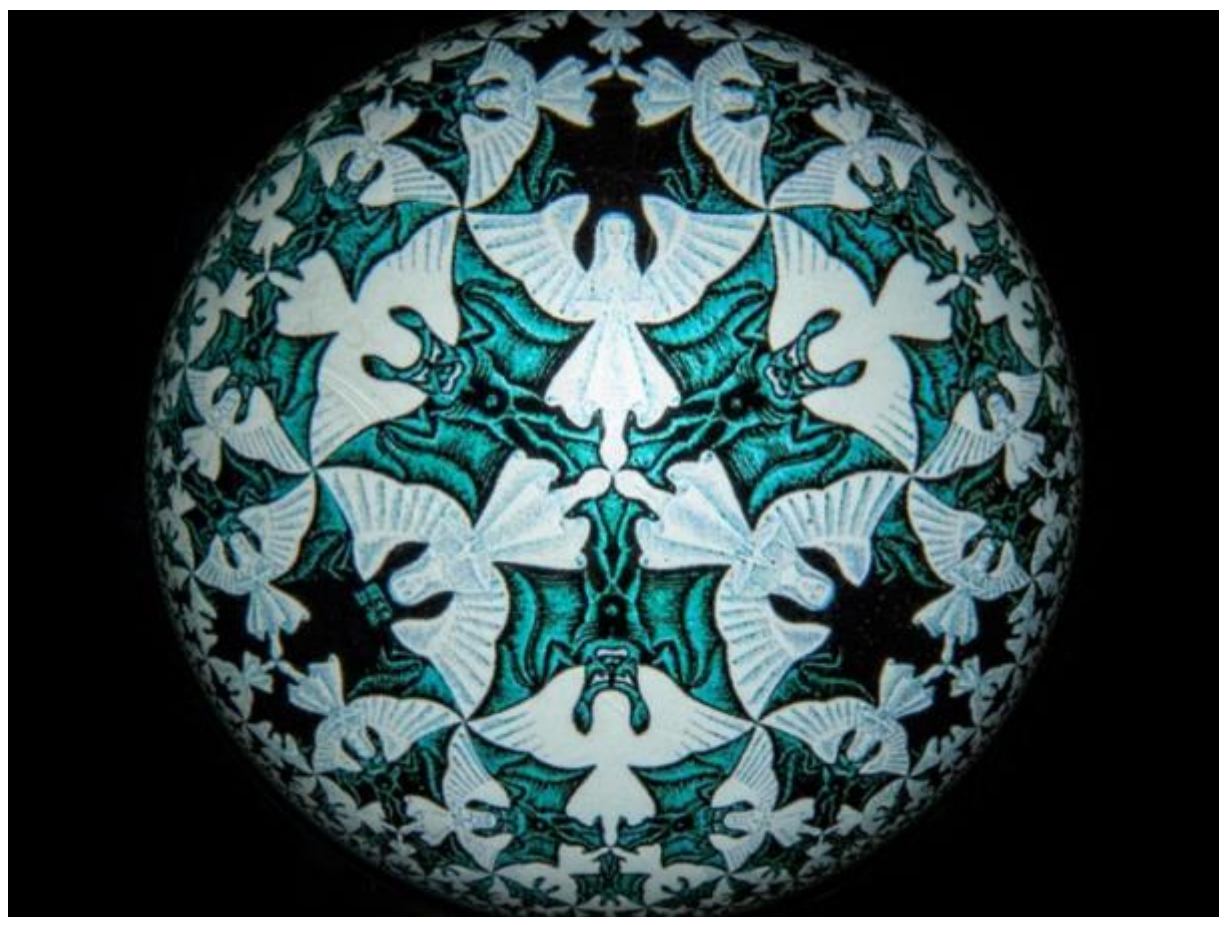

*Note*. Escher's painting "Circle Limit IV". From *Angels and Demons* by G. Pow (2018), flickr (https://www.flickr.com/photos/graeme_pow/45645163031). Licensed under **CC BY-NC-SA 2.0.**

Now, imagine being inside an AdS that had 3 dimensions of space, and with many soccer balls, each the same size, being located at various distances from you. Because of the negative curvature of ADS, the further away a soccer ball is from you, the smaller it would appear. This is similar to how, in Escher's 2-dimensional painting, *Circle Limit IV*, the angels and demons look smaller the further away from the center they are. But remember that each ball, regardless of its distance from you, would remain the same size—they would only appear to vary in size, owing to the curvature in space. In Maldacena's approach, we can imagine the outer surface of a 9-dimensional AdS being a 3-dimensional brane: as a visual illustration, picture the brane as being the surface of a sphere, with the AdS being the interior of the sphere.

In string theory, open strings must have at least one end attached to a brane. In Maldacena's approach, open strings have both ends attached to a brane. It is even possible to have multiple branes very closely stacked on top of each other. In this configuration, an open string can have one end on one brane and the other end on another brane (with both branes being in the stack of branes).

To eliminate gravity from the picture, the closed strings are all given very low energies. The lower the energy of each string, the less mass they have; and the less mass they have, the less gravity they exert on each other. In string theory, closed strings transmit the force of gravity and do not have to be attached to branes—they can move freely in whatever space there is, whether in branes or not. In Maldacena's approach, closed strings also have low energies and move in the AdS. Maldacena looked at how the stack of branes would behave if they were considered as a single object. It turns out that they exert a gravitational force on the AdS that they surround. Indeed, they act like a black hole: within the AdS, the gravity exerted by the brane (which, again, surrounds the AdS) gets stronger the closer an object gets to it. If you then were to place a closed string (a loop of string) with high energy very close to the brane and observed it from a distance, it would nonetheless look like a closed string with low energy—it would look as if its vibrations had slowed down significantly. Like a normal black hole, the gravitational pull of the stack of branes (now considered as a single object) causes the light emitted by the closed string to appear to a distant observer as if it had a long wavelength. And consequently, the light also appears to have low energy. But what's also important to note is that, because they are closed strings, they ensure the presence of gravity within the AdS.

The upshot of all of this is that we have a single physical situation—a stack of branes surrounding an AdS, like a metaphorical spherical shell enclosing its interior space—described by two equivalent descriptions. More specifically, the brane-centered explanation is synonymous with quantum field theory, which does not contain gravity; and the AdS-centered explanation is synonymous with string theory, which does contain gravity. Of course, AdS is not the kind of 3-dimensional universe we live in, nor, ultimately, the kind of 9-dimensional universe we live in if string theory is right. Nonetheless, the main takeaway message is rather striking: two radically different descriptions of the same physical situation are equivalent—*identical*—despite the fact that one contains gravity while the other doesn't, and despite the fact that they *each contain different numbers of spatial dimensions*. In physics, two or more descriptions or explanations that are equivalent to one another are known as *dualities*.

Work by Witten et al. (1998) made the link between Maldacena's discovery and holography more explicit. They provided a mathematical dictionary showing how a theory of an interior—an AdS—with any given number of dimensions is identical to a theory of its boundary—a stack of branes—with fewer dimensions. Thus, for example, a boundary explanation with 2 spatial dimensions can be identical to an interior explanation with 3 spatial dimensions. In that scenario, a black hole in a 3-dimensional AdS would be identical with a quantum field theory expressed in terms of a 2-dimensional boundary—a stack of 2-dimensional branes (Greene, 2011). But it should be stressed that two identical theories—one describing the boundary, and the other describing the interior—*look nothing at all like one another*. As we saw in our initial discussion of the holographic principle and black holes, the AdS and brane descriptions are as different from each other as a 2-dimensional hologram—with its squiggles etched into a plastic sheet—is from the 3-dimensional image it produces when illuminated by a laser. Nonetheless, the mathematical dictionary developed by Witten and his colleagues functions as the analogous laser does, translating the radically different theories into one another.

Greene (2000) discusses the duality that appeared in earlier work in string theory. For example, a cylindrical space with a certain radius and a certain configuration of strings wrapped around it generates the same physics as a cylindrical space with the reciprocal of that radius and a certain configuration of strings wrapped around it. Here’s another illustration using mirror symmetry: consider string theory’s six very small extra dimensions that are curled up within the large three dimensions with which we’re familiar. The shapes and sizes of these small dimensions—known as Calabi-Yau manifolds (see Figure 4)—play a central role in generating the masses and charges of particles, and so on. The extra dimensions can also have holes in them. But although the holes can be found in any of the extra dimensions, and although each dimension can have any number of holes, it is only the total number of holes that makes a difference to the physics that the holes give rise to. So, holes residing in different topologies can have the same physical effects so long as their total number is the same (here a topology is a space made of a certain shape and size of the small extra dimensions).

**Figure 4**
Calabi-Yau Manifold

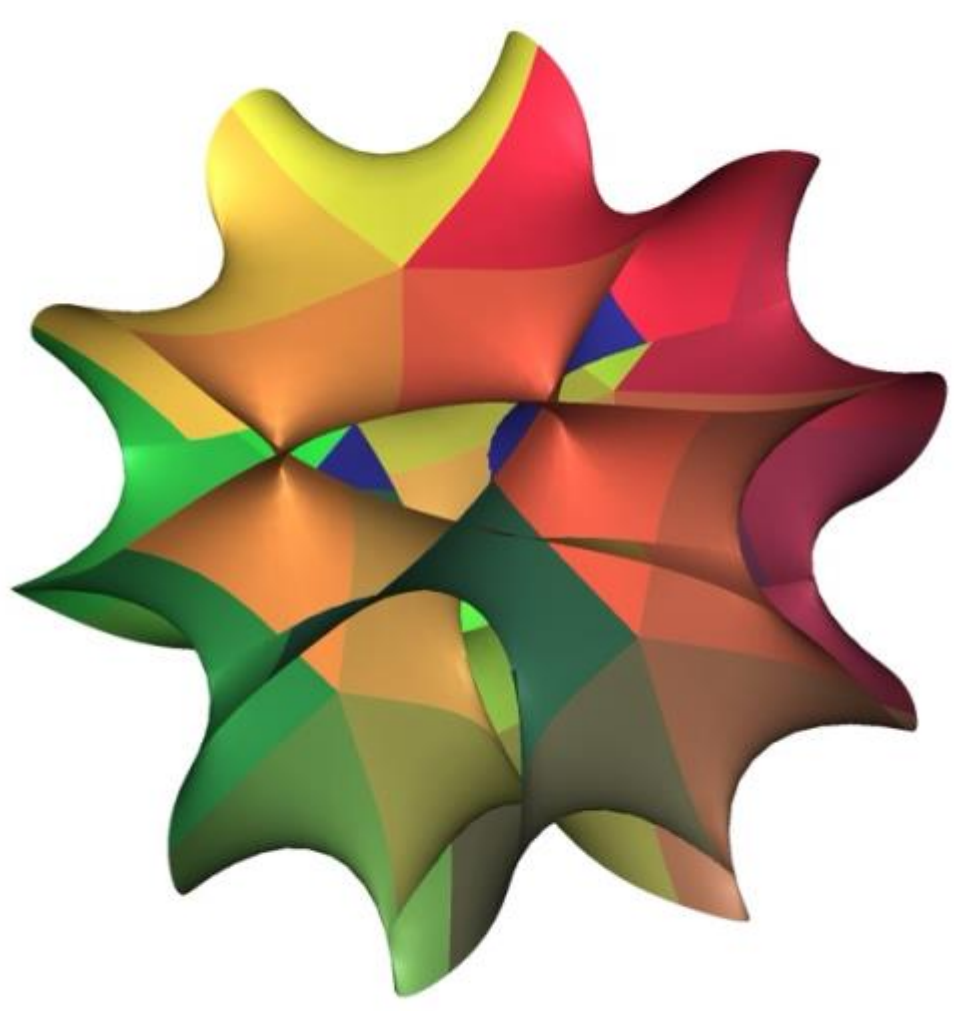

*Note*. An artistic depiction of a Calabi-Yau manifold. From *CalabiYau5* by A. J. Hanson (2014), Wikimedia Commons (https://commons.wikimedia.org/wiki/File:CalabiYau5.jpg). Licensed under **CC BY-SA 3.0.**

Here’s one final example that encapsulates the last two examples of cylinders and Calabi-Yau manifolds: String theory at one point came in five different versions, known as Type I, Type IIA, Type IIB, Heterotic O, and Heterotic E. But it eventually became apparent that all five were the same theory just in different forms; and the framework for making these interconnections explicit is M-theory (see Figure 5). Indeed, an earlier approach to merging quantum mechanics with general relativity, known as supergravity, also approximates string theory. Whereas the supergravity approach is a point-particle quantum field theory in nine spatial dimensions, the original five string theories contain extended strings as their primary entities—and those strings move in the same nine dimensions of space (a version of supergravity with ten dimensions of space, in turn, approximates M-theory). The idea here is that because strings are so incredibly tiny, they look and behave like point-particles from “far away” (at large-scale distance). But just as importantly, even though, say, the Type IIA and Type IIB string theories are mathematically equivalent and describe the same physical consequences, the geometry of space is different in each theory. So, in these other examples from string theory, we again see the theme of duality: *different geometries or perspectives giving rise to the same physics*.

**Figure 5**
*M-Theory*

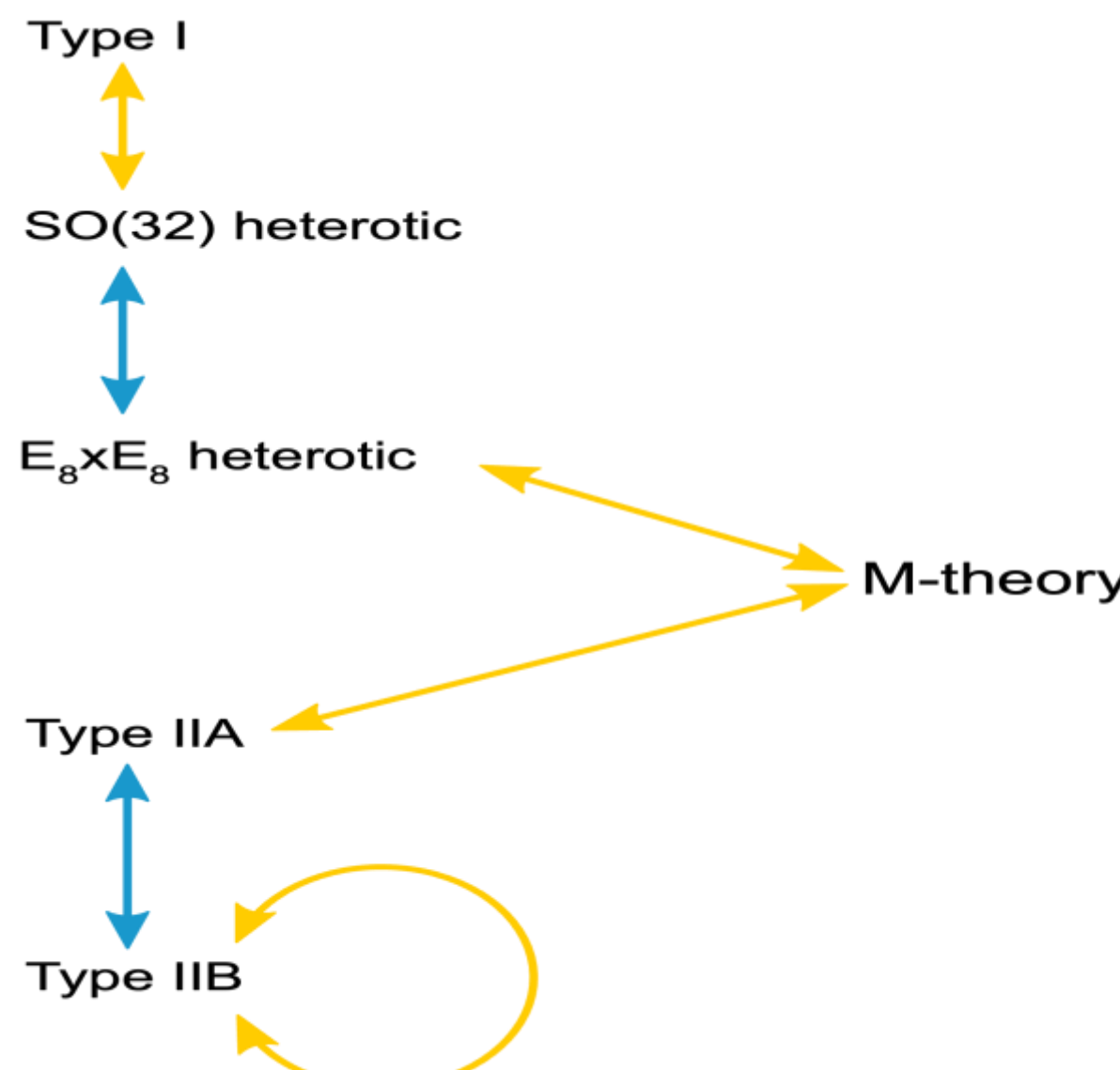

*Note*. The equivalent theories of M-Theory. Adapted from *StringTheoryDualities* by A. Dunkel (2021), Wikimedia Commons (https://commons.wikimedia.org/wiki/File:StringTheoryDualities.svg). Licensed under **CC BY-SA 4.0**.

Philosophers will find the basic point here quite familiar: it's one manifestation of the under-determination thesis of Quine and Duhem, according to which an observation can, in principle, be equally-well explained by two or more explanations that make the same empirical predictions. (To arbitrate between the possible explanations, it is typically necessary to come up with at least one kind of test that can decisively distinguish between them empirically.) And ever since work by Poincaré (1952), physicists and philosophers of physics have known about the theoretical possibility that different spatial geometries can in principle be consistent with observations.

## Duality and the ITP

One relevant feature of string-theory duality for the ITP is that it makes certain calculations much easier than they would otherwise be. In one case, a team of string theorists actually found the solution for a very complex geometric problem faster and more efficiently than a team of mathematicians (it was a contest of sorts) (Greene, 2000). This shows that different but equivalent ways of representing reality can make tasks easier to complete. If the ITP is true, then it *parallels the way that duality in string theory makes certain calculations much easier than they otherwise would be*. Reality, whatever it is like in and of itself, may be very complex (in some sense), but our perceptions might have evolved to represent that reality in a much more streamlined and simplified way. Again, the ITP argues that our perceptions evolved not to represent reality, but rather to enhance reproductive fitness. More generally, however, while M-theory may not, as I've said, turn out to be the correct theory of quantum gravity, its dualities—including Maldacena's string-version of the holographic principle—nonetheless show that it is at least possible in principle to accurately represent phenomena in more than one way.

The ITP, however, argues that perception does not even have to represent reality accurately; it only needs to promote reproductive fitness. So, the bar that perception needs to surmount is probably much lower than what dualities in physics must surmount: dualities in physics need to be very accurate if they are to be strictly equivalent to one another, whereas perception, according to the ITP, only needs to promote reproductive fitness. And if the ITP is right, the objects and 3-dimensional space that physics posits to exist are only perceptual representations; they do not exist in the reality external to our perceptions. Thus, particles, stars, planets, and even strings, and so on, would merely be representations—useful fictions—with no literal existence outside of perception.

And there are other reasons for thinking that we evolved to perceive external reality in terms of space. Work by physicists suggest that our spatial perceptions indeed function like an error-correcting code (recall the earlier discussion of error-correction). For instance, Almheiri et al. (2015) show that 3-dimensional space can have an error-correcting function that tamps down on noisiness in data. And Pastawski et al. (2015) show that spacetime can function as a form of error-correction for quantum-mechanical data.

## But what is space?

Let's set our discussion of the holographic principle aside for the time being. Earlier I pointed out that physics is still trying to reconcile the incompatibilities between its two main foundations, general relativity—which essentially explains the physics of the very large—and quantum mechanics—which essentially explains the physics of the very small. String theory is widely regarded as one of the most promising, if not the most promising, approaches to reconciling the two, and the most well-developed approach to boot. An alternative approach, however, is known as loop quantum gravity. One of the main selling points of the loop approach is that it aims, right out of the gate, to be what's known as a "background-independent" framework for merging quantum mechanics with general relativity. What this means is that it does not begin by presupposing the existence of space (and time); the approach aims instead to explain not only how space behaves, but how it emerges in the first place. By contrast, string theory is often alleged to be the opposite, namely a "background-dependent" framework that presupposes the existence of space, and from that starting point aims to unify quantum mechanics with general relativity. On the other hand, the holographic principle suggests that string theory is also moving in the direction of being a background-independent approach that explains the emergence of space. Nonetheless, the approach of loop quantum gravity is another direct way of thinking about what space ultimately is.

A good entryway for understanding the essence of the loop approach is Rovelli's (2018) discussion, which I'll draw from.[5] First, one of the foundational premises of the loop approach is one that we have already seen with black holes and the holographic principle: that space is not a continuum that is infinitely divisible, but rather that it hits rock bottom at some point, namely at the Planck-length. In classical physics (i.e., Newtonian physics), including general relativity, space is treated as a continuum: take any region of space, and you can, according to the view of space as a continuum, continuously subdivide it, endlessly. For example, a piece of space 10 cm wide could be divided into 2 pieces that are each 5 cm. One of those pieces could then be divided further, and so on, without end, ad infinitum. When trying to merge quantum mechanics with general relativity, however, the continuum of the gravitational field, which makes up space,

[5] Smolin (2008) gives an historical overview of the development of loop quantum gravity.

presents problems, because it too is a continuum.

Rovelli draws an analogy between this problem and Zeno's well-known paradox of Achilles and the tortoise, who are having a race with each other. In the paradox, Achilles is (supposedly) impossibly tasked with covering an infinite distance between himself and the tortoise. In virtue of its slowness, the tortoise is also given a head start when the race begins. Since Achilles can run faster than the tortoise, he should be able to overtake the latter. But Zeno argues that by the time Achilles has reached the place where the tortoise started the race, the tortoise will have since moved a bit ahead of that place. And when Achilles then reaches that next point, the tortoise has since moved a tiny bit ahead of that point. Thus, according to Zeno, each time Achilles manages to reach the point where the tortoise was just prior, the tortoise will always have advanced just a little bit more beyond that point. Zeno's implication is that this should continue infinitely—and hence Achilles should never quite catch the tortoise. With space, the question is whether it truly is a continuum that can be infinitely subdivided—just as, according to Zeno, the gap separating Achilles and the tortoise can always be subdivided without end.

When quantum mechanics is applied to the continuum of the gravitational field, space warps and bends chaotically at very small scales—and it continually does so at increasingly small scales, again without end. This essentially results in calculations that yield infinities and thus are nonsensical. But with the loop approach, just like the holographic principle and string theory more generally, space has a granular structure—space is discrete—and there is a basement level at which it no longer can be subdivided. According to the loop approach, space is voxelated—it is made up of very tiny cubes. (And since space is discrete, volume and area are also discrete—they come in packets—as can be shown by Paul Dirac's general equation of quantum mechanics.) As in string theory, the discreteness of space in the loop approach tames the infinities that otherwise emerge if space is a continuum that's infinitely divisible.

One of the physicists to first think about how general relativity and quantum mechanics might be reconciled was John Wheeler. According to Wheeler, once you got down to the Planck-length, the nature of quantum fluctuations would become so severe and tempestuous that the gravitational field, which comprises space, would resemble a jumbled, foamy-like structure. In a sense, this idea of a "quantum foam" is meant to visually-represent the range of different geometries that space can take at and near the Planck-length, owing to the probabilistic nature of quantum processes. Another way to look at this is to think of a wide range of different spatial geometries superimposed onto one another at the tiny scales found at and near the Planck-length. An initial mathematical approach to grasping how quantum uncertainty (quantum fluctuations) would apply to space at that scale was developed by Wheeler and Bryce DeWitt. This was one of the main launchpads for the loop approach, which made Wheeler and DeWitt's first foray more rigorous.

So how does loop quantum gravity try to explain the emergence of space? It models space and its emergence using spin networks, which are essentially graphs that depict the elemental pieces of space and their properties (see Figure 6). Like any basic network, the two fundamental ingredients of spin networks are nodes and links. Nodes are dots, and links connecting them are lines. Each node in a network represents a volume of space, which, again, is granular. And each link in a network represents an area of space, which is also granular. Each node and link can have a discrete value, although there is also a minimum size for each, which is the Planck-length. Roughly speaking, the way that nodes are connected by links can describe the curvature of space: links are assigned "spin values", which describe the curvature of the gravitational field.

**Figure 6**
A Spin Network

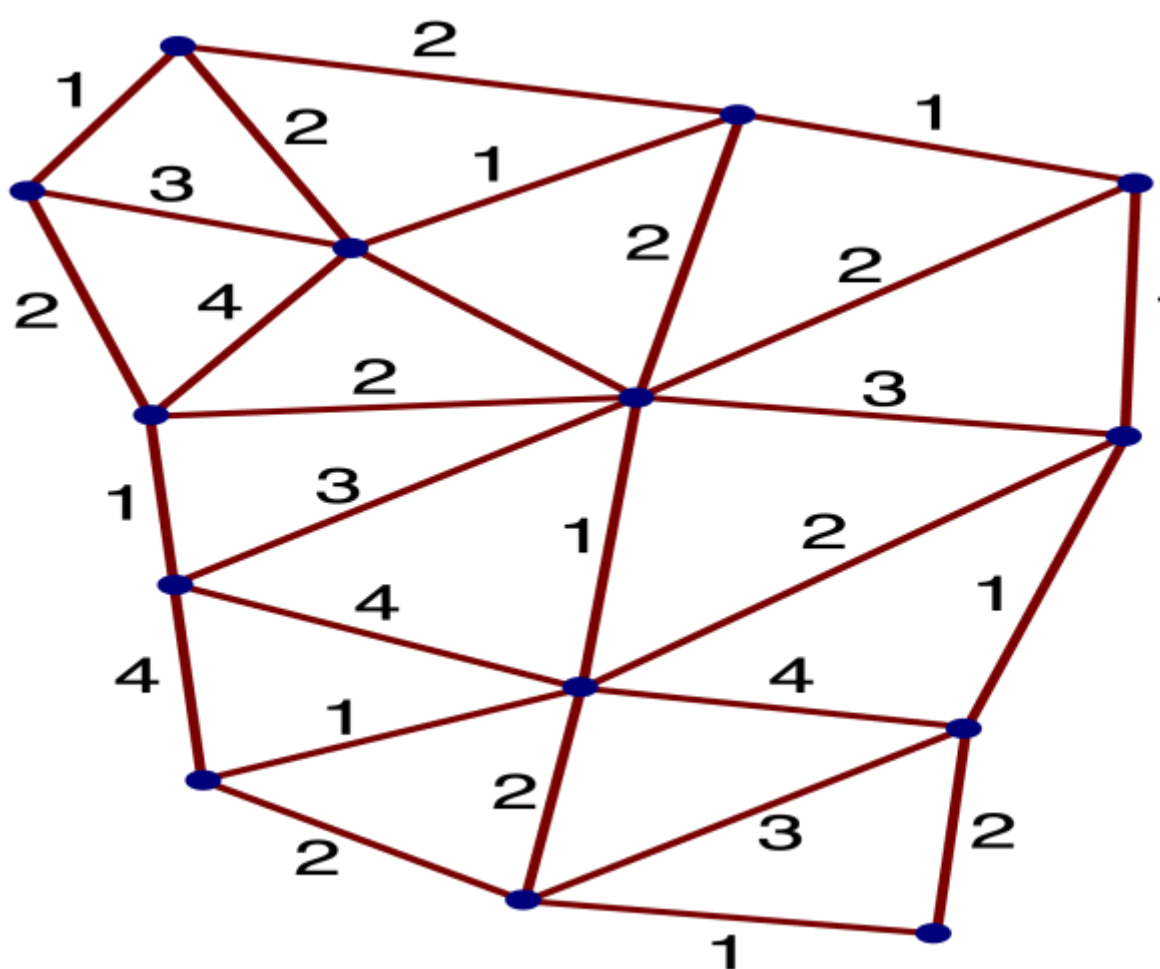

*Note*. In loop quantum gravity, a spin diagram depicts a spin network. Nodes are dots, and links are lines connecting various dots. Each link has a spin value, which describes the curvature of the gravitational field. From *Spin network* by M. Poessel (2008), Wikimedia Commons (https://commons.wikimedia.org/wiki/File:Spin_network.svg). Licensed under **CC BY-SA 3.0**.

To make the discussion more concrete, we can visualize spin networks as representing the "atoms" of space. Spin networks do *not* describe elements *within* a pre-existing spacetime. Spin networks describe the *individual elements of space*—quanta of space—which, through their relation to one another, make space. But to call these individual elements of space, space, is something of a misnomer, since no individual element of space is space per se. They are more like individual threads which, when interweaved with one another, build the spatial fabric.[6] Moreover, we shouldn't lose sight of the fact that these networks of elements, which collectively describe space, are fundamentally quantum in nature. Hence, Wheeler's notion of a space foam: the cloud of different geometries superimposed on one another and described by quantum probabilities.

According to loop quantum gravity, nature is fundamentally made of quantum fields (or covariant quantum fields, to be more precise). Quantum fields are fundamentally embedded in and built out of the spatial fabric, which itself is also a quantum field.[7] Quantum fields can be likened to fluid-like or spindle-like substances suffusing space, and whose properties are intrinsically quantum-mechanical, as their name indicates. Finally, "spinfoams" are spin networks that evolve in time: they describe how the geometry of space (the gravitational field) at or near the Planck-length evolves over time, given quantum uncertainty.

[6] Another alternative approach to explaining quantum gravity is known as "causal-set theory" (Sorkin, 1991). To very briefly summarize, according to causal-set theory, nature is fundamentally described by discrete events which are causally related to each other in a network. From these elementary events and their relations, all the rest of nature emerges. Smolin (2019) has also developed a version of causal-set theory.

[7] Rovelli (2018) likens quantum fields to Anaximander's *apeiron*: the fundamental substance from which everything in the universe is formed. Physicists often point out that particles are not actually "real": they are a kind of shorthand way of conceptualizing the behavior of fields—and ultimately quantum fields—which make it easier to grasp the underlying phenomena. But what we call particles are really localized disturbances in fields, and they of course can move and can interact with each other (e.g., Carroll, 2020; Lange, 2002; Rovelli, 2018).

Let's close by looking at one final approach that differs from string theory and loop quantum gravity in an important way. Whereas the string and loop approaches attempt to take gravity and "quantize" it, another approach tries to go from the bottom-up: it starts with quantum mechanics and tries to find gravity, and thus space, within it. So, it might be possible to explain how space emerges by using the wave-function of quantum mechanics and the appropriate version of the Schrodinger equation. This would be a quantum-first approach to explaining the emergence of space (Carroll, 2020). One of the basic ingredients of this quantum-first framework is quantum entanglement, particularly the entanglement found in quantum field theory. Since space is supposed to emerge, the quantum-first approach is similar to the loop approach in that it doesn't have the luxury of starting with a spatial geometry; space must instead pull itself up by its bootstraps using only the abstract ingredients of quantum mechanics. The quantum-first approach aims to show how an empty space with no matter or energy in it can emerge. The empty units of space are referred to as "degrees of freedom". As in the loop approach, each degree of freedom is like a thread, and collectively the threads are woven together to form space. In this way, space emerges. Degrees of freedom are not highly entangled with each other merely by being nearby to each other; rather, degrees of freedom are nearby to each other *if* they are highly entangled with each other.

In the quantum-first approach, it's argued that our actual world has a specific kind of Hamiltonian (which is essentially a compact description of the total state of energy in a system). This view entails that degrees of freedom are entangled in such a way that they produce the kind of space that we're accustomed to, where local interactions are the norm. Furthermore, the quantum entanglement between degrees of freedom is similar to entropy; and the entropy of a set of entangled degrees of freedom is equivalent to its area. Carroll (2020) sums it up this way:

> … mathematicians long ago figured out that knowing the area of every possible surface in a region is enough to fully determine the geometry of that region; it's completely equivalent to knowing the metric everywhere. In other words, the combination of (1) knowing how our degrees of freedom are entangled, and (2) postulating that the entropy of any collection of degrees of freedom defines an area of the boundary around that collection, suffices to fully determine the geometry of our emergent space. (p. 284)

Now let's reiterate the central point, lest it's been lost in the details. According to the loop and quantum-first approaches to quantum gravity, *space is not fundamental—it emerges*. In this sense, the fundamental premise of these approaches is quite arguably even more radical than the holographic principle. Equally radical is that, according to the loop approach, time, just like in general relativity, is part and parcel of space. So, the upshot is that time also emerges with space—which means that both space and time (space-time) are not fundamental.

## Is time fundamental?

We've looked at the holographic principle and space, but we have only just mentioned time. So, let us know now ask where time fits into all of this. And let us begin by looking at the question in terms of Einstein's relativity, and then from the point of view of quantum mechanics.

A well-known feature of special relativity is that simultaneity is relative and depends on a reference frame—for instance, it depends on how one observer is moving relative to another. And the farther apart that two observers in space are, the larger the effect of relative motion between them will be. For example, imagine two people: Jerry, located somewhere on Earth, and Kramer, located on a planet far away, some 10 billion light-years from here.[8] Because of the vast distance separating the two, any difference in motion between them will amplify the difference in how they register the same events. If we assume that both Jerry and Kramer are seated and moving at precisely the same speed (all things considered), they will both register all events in the universe as occurring at the same time. But if Kramer started moving away from still-seated Jerry at 10 miles per hour, then events that would be registered by Jerry as happening right now would for Kramer be registered as happening 150 years ago according to Jerry. Conversely, if Kramer started moving towards Jerry at 10 miles per hour, then suddenly the events that would be registered by Jerry as happening right now would for Kramer be registered as happening 150 years into the future according to Jerry. Again, the reason why the two observers arrange the sequence of events so differently is because the vast distance separating the two observers amplifies the small difference in motion between them.

Two observers in relatively close proximity must be moving at very different speeds for the two to noticeably register the sequence of events in a different way. But large distances separating two observers can amplify the difference in their speeds and create the same effect that a large difference in speed would: either way, they are led to register the sequence of events differently. We can generalize this by applying it to every point in spacetime: any difference in motion between hypothetical observers at every place in spacetime will yield differences in how they register the sequence of events. The upshot is that, from the point of view of special relativity, the passage of time is merely apparent—an illusion—and all events, however they are registered by observers, are frozen in place, neither coming into being nor ceasing, but rather always in existence. It should go without saying, but the view from the physics of relativity is drastically different than how we experience the world. Somehow, the flow of time is an illusion produced by our conscious experience—it is merely apparent, but not a real, fundamental feature of physical reality. We experience an objective passage of time from our subjective first-person point-of-view, but there is no such objective passage of time according to the physics of special relativity. In philosophy, this Einsteinian perspective on time is known as eternalism, or the "block universe" (as depicted in Figure 7).

[8] This discussion is adapted from Greene (2004).

**Figure 7**
The Block Universe

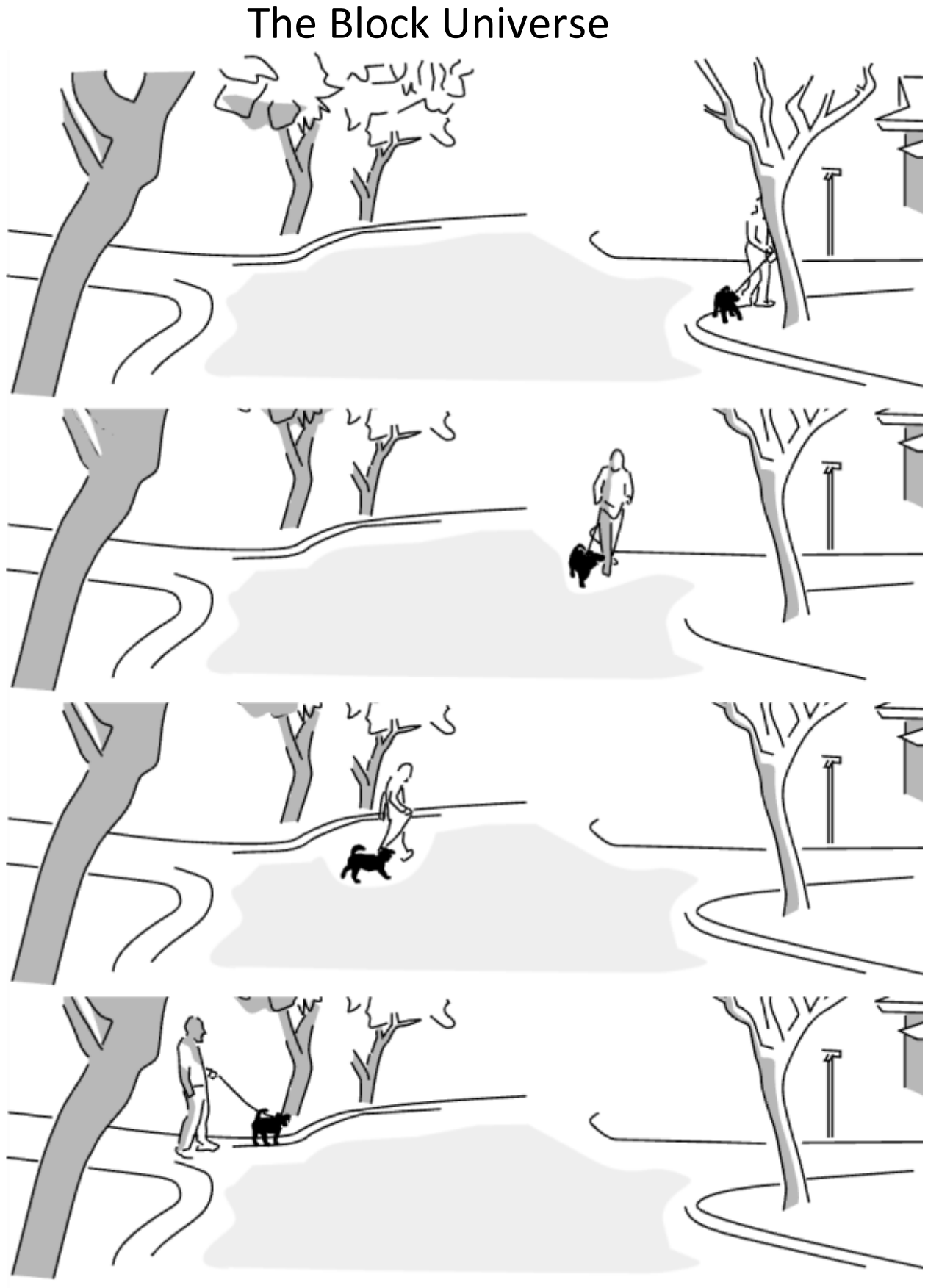

*Note*. The block universe is a view that postulates that the flow of is an illusion—that time is not real, merely apparent to observers such as us. Instead, every fundamental moment of time is actually a frozen slice that exists eternally. From *Illustration of block universe* by B. Crowell (2021), Wikimedia Commons (https://commons.wikimedia.org/wiki/File:Illustration_of_block_universe.svg). Licensed under **CC BY-SA 4.0**.

Now let's examine time from a quantum perspective, focusing in particular on the Wheeler-DeWitt equation. The Wheeler-DeWitt equation is an attempt to apply a single quantum wave-function to the entire universe. Page and Wootters (1983), for example, illustrate how this might be done. Imagine taking snapshots of the entire universe at a regular interval—say, every Planck time, which is the time needed for light to traverse one Planck length. Suppose further that we associate a time with every snapshot, and every snapshot includes a clock that displays the time at which the snapshot was taken. We can then give every snapshot its own wave-function. And because we're dealing with the Wheeler-DeWitt equation, which is a wave-function for the entire universe, we can take each snapshot's wave-function and combine them into a single wave-function: a superposition of each of those individual wave-functions. What we get, then, is one wave-function that describes the physical state of the universe, at all times and all at once. In other words, according to the Wheeler-DeWitt equation, time is not fundamental. Rather, it "emerges", somehow.

But there are really two possibilities here, as Carroll (2020) points out. If quantum gravity turns out to obey a formulation of the Schrödinger equation, then almost every quantum state will contain time that runs infinitely into both the past and future. In which case, there would have been time before the Big Bang too. The other possibility is that, as is the case in the Wheeler-DeWitt equation, time drops out of the picture entirely. This would be true if the universe has a total of zero energy. But the possibility that the universe's total energy is precisely zero is not as far-fetched as it might seem, for the simple reason that all energy—and thus all matter—produces gravity, which produces negative energy. And negative energy cancels out positive energy.

So, if the universe is "closed", such that its geometry closes in on itself like a sphere or torus, then, mathematically, the universe's total positive energy will be precisely canceled out by the negative energy of gravity. The universe would thus have a total energy of zero (Vilenkin, 2007). Greene (2011) points out that according to inflation, which is the dominant paradigm of current cosmology, the universe would look like an expanding bubble to a hypothetical observer standing "outside" of the universe; which is to say that it would have a "closed" geometry, like a sphere, but would also look to the observer like it exists infinitely far into the future. But for observers inside the universe, space would look infinite in extent, while the universe itself would appear to have existed for only a finite time. So, for the outside observer, space would look finite and time would look infinite; and for the observer inside, space would look infinite and time would look finite. Thus, it may be that the universe does have zero total energy—from the point of view of the hypothetical observer outside of the universe—but it may look otherwise to observers on the inside. This is because an observer outside the universe would see it as finite; its total positive energy would be exactly canceled by the negative energy associated with gravity.

One way to visualize the passage of time is use the analogy of a film projector that projects each frame onto a screen, frame-by-frame. Each frame represents the current instant—the moment right "now". And if the block universe-view is correct, then each moment depicted in a frame is a frozen state. If we were to look at the universe from the outside, we would see each and every moment of time as a frame; and each and every moment would just exist, neither coming into being nor disappearing. We would see a collection of frozen film frames with no temporal change contained within any of them; and none of the frozen frames would go out of existence and be replaced by a new frame. There would be no flow of time, in other words, either within a frame or occurring between frames. To be clear, there would simply be a collection of frozen frames and no passage of time.

One possible way there might be a real flow of time would be to claim that there is real change, and that the fundamental unit of change occurs during the smallest hypothesized unit of time in physics, namely Planck-time. (The Planck-time, recall, is the time light takes to traverse one Planck-length). If this is a possibility, then we're left with the following options. A "quantum" of time, with a duration equivalent to the Planck-time, could underlie a real passage of time; and this would be a mechanism for producing real rather than apparent change, and thus there would be a real flow of time. Or, if the Wheeler-DeWitt equation is the right way to think of the universe quantum-mechanically, and if the universe has a total energy of zero, then time may be an illusion—unless it can emerge in some other unspecified way. On the other hand, the conception of spacetime given by special relativity may be retained within a new quantum account of time, in which case the passage of time would be merely apparent rather than real, as implied by the block universe-view.

But if, from the point of view of physics, the passage of time is indeed not real, we are still left with our subjective experience of time. One plausible outline of how we can experience the flow of time even if it's not real is given by Barbour (2001), who also argues that time is unreal. Suppose a cat jumps onto a table and the brain captures the cat's jump as a collection of moments which are consciously experienced. If the passage of time is an illusion, the state of the brain that realizes the moment we experience "now" would have been affected by the state of the brain that realized the moment we experienced just prior; and the brain state that realizes the moment we experience next—the next "now"—would likewise be affected by the brain state that realizes our experience "now". In this sense, physical relations between states of the environment also affect brain states: we perceive changes to the environment, and these changes inform each of our conscious experiences, via different states of the brain. Each moment of conscious experience that seems to be experienced as a moment in the flow of time is realized by a static brain state. Because, if the block universe-view is correct, then each moment of conscious experience can only be realized by a frozen state of the brain. Because, as per the block universe, there is no flow of time either within frames of time or between them. A conscious experience at one moment would seem to us as if it's continuous with what was experienced in the prior moment. Hence why, for any particular conscious state at a given moment, it would seem as if time were flowing normally, from the past towards the future. We would subjectively experience a continuous flow of time that stretches from the moment just prior, to the moment now, and towards the next moment. (Intriguingly, people with certain brain abnormalities might report atypical and uncanny experiences of the flow of time, [e.g., McGilchrist, 2021]).

This is obviously just a simplified sketch of how the process might occur in the brain. The "color phi phenomenon" is another analogy that can illustrate how static states of the environment could be perceived as a continuous flow. In the color phi phenomenon (see Figure 8), a sequence of colored dots flashes: first a blue circle is flashed on the left side, and then, relatively rapidly, at some distance to the right, a red circle is flashed. But subjectively, we experience the blue dot moving to the right and transitioning to the red dot midway. Of course, the motion of the dot and the transition between colors midway are only apparent, not real, and the brain detects both the blue circle and red circle *before* presenting an interpretation of the events in our consciousness: the brain first detects the blue dot, then (rapidly) detects the red dot, and then creates a perceptual experience of the blue dot moving to the right and changing to a red dot, midway through the trajectory. In other words, the second dot—the red dot—is detected before the perception of motion and transition are experienced; the latter moment informs the prior moments.

**Figure 8**
The Color Phi Phenomenon

**What is Actually Shown**

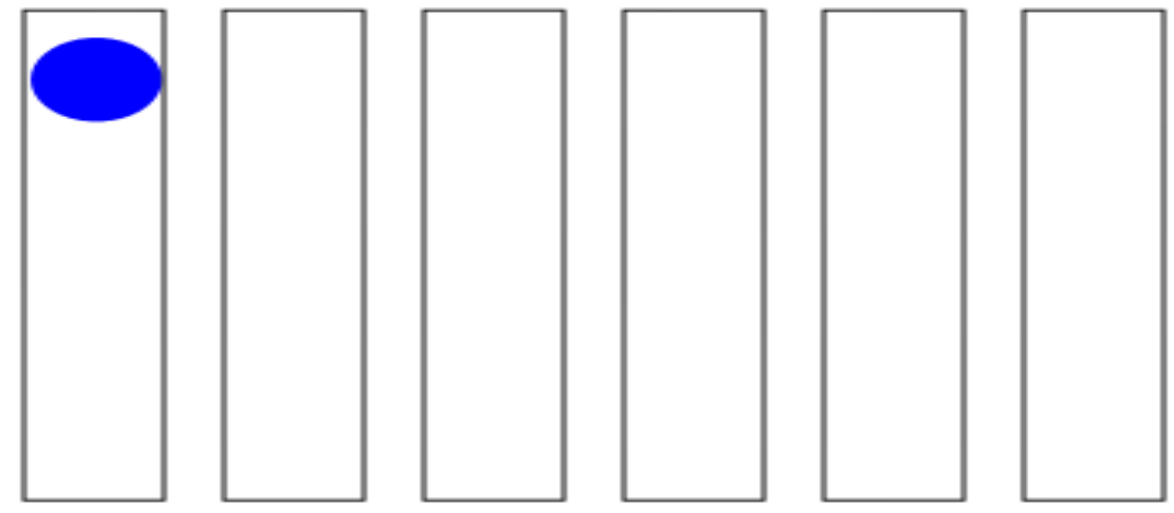

Time

**What Subjects Report**

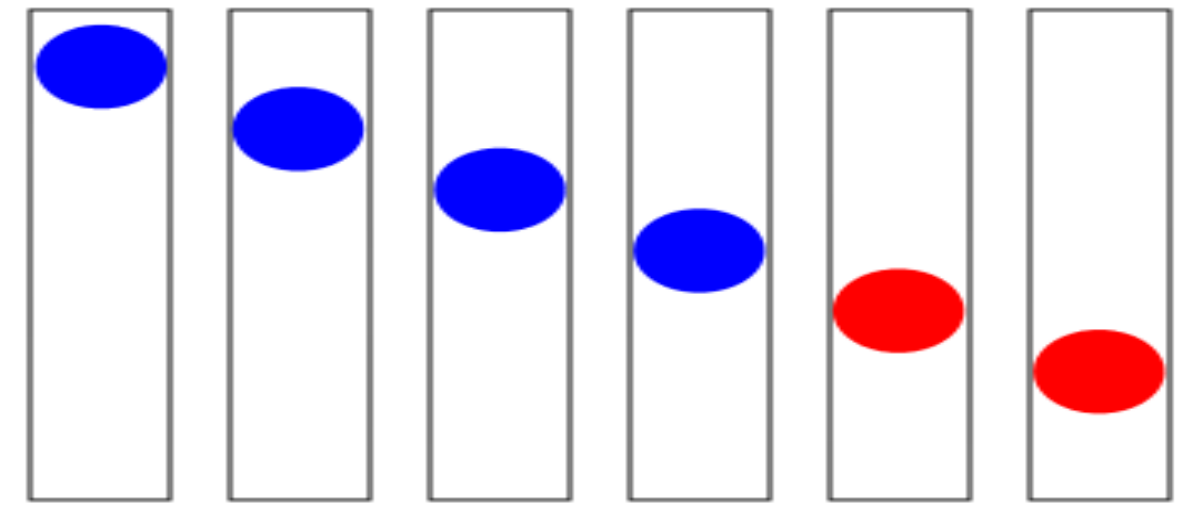

Time

*Note*. In the color phi phenomenon, a blue dot on the far left is flashed, followed in close succession by the flashing of the red dot on the far right. People report seeing the blue dot move towards the right and turn into the red dot midway through its trajectory. This demonstrates that apparent motion can be generated by the brain—there need not be any such motion in the world. Adapted from *Color phi phenomenon* by Alecmconroy (2006), Wikipedia (https://en.wikipedia.org/wiki/Color_phi_phenomenon). Licensed under **CC BY-SA 3.0**.

Conscious experience of the color phi phenomenon could just be a series of brain states that are timeless, as per the block universe. If the flow of time is an illusion, then presumably there would also have been a kind of selection pressure on animal cognition and on consciousness to perceive the world in a way in which time appeared to flow. Although, admittedly, you do need to rewire your brain a bit to really grok how natural selection—which is a process—works in a reality sans time. To reiterate: If the passage of time is an illusion, prior moments would affect the state of the brain "now", which in turn would affect the state of the brain at the next moment. A brain state at one moment would seem as if its continuous with what came before in the prior moment. In the case of the color phi phenomenon, the illusion of apparent motion and color transition mid-way would be produced only after both the blue and red dots were perceived.

Given our discussion of time, it is also fairly easy to see that different moments of time are related to each other via the wave-function of the universe given by Wheeler and DeWitt; indeed, the coherency between moments of time can be explained in terms of entropy: the apparent flow of time is a manifestation of the second law of thermodynamics, whereby, probabilistically, more ordered states evolve into less ordered ones.

## What might external reality be like?

It is worth considering a counter-argument to Hoffman's theory at this point. The philosopher David Chalmers (2022b) has this to say about it:

> My former Ph.D. student Brad Thompson devised a thought experiment about Doubled Earth, where everything is twice as large as on Earth. When Brad sees an object one meter tall, Doubled Brad sees an object that we would say is two meters tall. But Doubled Brad is isomorphic to Brad, so when he sees this object he has the same sort of experience that Brad has when seeing a one-meter-tall object. Is Doubled Brad suffering from an illusion, seeing the object as half as

> large as it actually is? Perhaps Hoffman will say yes. On an Edenic model with absolute sizes, one would say yes. But most people find it far more intuitive to say that both Brad and Doubled Brad are correctly perceiving the world. And certainly this is what spatial functionalism will predict. Once again, Hoffman's assumption that there's only one objectively correct mapping from world to mind seems implausible, at least after we have fallen from Eden. Now, Hoffman might argue at this point that we have given up on objective reality. I don't agree—we have merely given up on the Edenic model of objective reality. There is still an objective world of structures out there. Objects in the world really do have colors and sizes. It's just that what makes them the colors and sizes they are is partly the roles they play and not some absolute, intrinsic nature. (p. 35)

Contrary to Chalmers, I think this is missing the point and I think Doubled Brad *would* be experiencing a kind of illusion. Let's assume that Doubled Brad is a member of Doubled Humans, who evolved to see the way they do (though we could just as well assume that they see as they do as a matter of brute fact, for the purposes of the thought experiment). Presumably, seeing everything as half the size it actually is would nonetheless allow Doubled Brad (and Doubled Humans in general) to get by on Doubled Earth just as well as Brad's vision and ours let us get by here on Earth. Still, that doesn't mean Doubled Brad is perceiving (external) reality as it actually is. His visual perception, although it works well for him pragmatically speaking, is still producing a kind of illusion. When we ask whether we're perceiving reality as it is—in this case, whether we're perceiving size and space as they actually are—we don't just want to know whether we're able to successfully navigate reality as we perceive it; we want to know whether we're accurately perceiving reality…as it actually is, in and of itself. In the case of the thought experiment presented by Chalmers, Doubled Brad is not accurately perceiving size and space as they are in and of themselves, yet, on Doubled Earth, as on Earth, there are still nonetheless size and space itself. As we've seen, though, given Hoffman's argument, as well as the case from physics, there is some reason to think that maybe even size and space do not exist at all in external reality. A similar rejoinder could be made to Chalmers' reply about colors: If Hoffman's theory is correct, then whether or not colors are playing a functional role in our perception, so long as colors—namely their intrinsic qualities, as they seem in our subjective experience—are not actually out there in the external world, then we would in fact be experiencing an illusion. (And anyway, even irrespective of Hoffman's theory, since there are no colors in the external world [no intrinsic colorness, in the way they seem in our subjective experience], our experience of color is still a kind of illusion).

Chalmers (2022b) also presents another counter-argument to Hoffman's theory:

> Hoffman might try to argue that we can't even know about structure. Maybe he could run his rewiring argument for numbers: When there are two balls in the world, we could experience three balls, and vice versa. But it's easy to see this can't

> work. We'll remove one ball from what looks like a pile of two balls and it will look like there are now three balls! So Hoffman's arguments pose no obstacle to knowing about structural aspects of the world. (p. 35)

I don't think this is a good counter-argument to Hoffman's theory either. For one thing, what makes Chalmers think that if we were to remove one ball from a pile of two balls that we wouldn't then experience two balls? Similarly, what would rule out the possibility that if we remove the last ball we would nonetheless experience one ball? It may seem counter-intuitive, but according to Hoffman's theory, there is no reason whatsoever to think that the perception of something like a ball must in any way follow the intuitive way that balls are related to each other in Chalmers' example. Indeed, I don't think intuition is a good guide for evaluating Hoffman's theory. After all, Hoffman's theory says that we evolved a user interface that represents reality in a radically different way than what reality actually is like in and of itself (whatever it is like). More to the point, according to Hoffman's theory, objects do not exist in external reality in anything like the way we perceive them, and nor does 3-dimensional space exist in reality either. So, if we see and interact with balls, there is nothing in external reality that must necessarily look and behave ball-like. According to Hoffman's theory, whatever role that ball perceptions are playing for us, they are playing a fitness-seeking role, not a role that represents reality as it actually is. Given his focus on structure, however, perhaps a variant of Chalmers' argument could work if, instead of saying that we perceive the structure of reality per se, he said we perceive the *fitness-relevant* structure of reality (and what's fitness-relevant is what's relevant to the fitness of a particular organism or species). But this is what Hoffman's theory is saying anyway; and Hoffman's theory says that what's fitness-relevant about reality will look nothing at all like what reality is like in and of itself.

We can now ask the question of what external reality might actually be like if the ITP is true. According to the holographic principle, the three-dimensional space that we perceive is a kind of illusion: it is the "projection" of a two-dimensional surface that surrounds the three-dimensional volume that we inhabit. Oddly, though, there is nothing in the holographic principle that explicitly says there is any such projection happening. So perhaps the most straightforward reading of it is to say that the three-dimensional space that we perceive is not actually what external reality is like. It might simply be—and in accordance with the ITP—that the three-dimensions that we perceive are a data-compressing and error-correcting interface that we evolved to perceive fitness, not external reality in and of itself. Physics, thus, may have inadvertently discovered the same conclusion as the ITP. And if so, then we actually live not in three-dimensions, but rather on a two-dimensional surface.

On the other hand, the other main approaches in quantum gravity strongly suggest that space "emerges" from the stitching together of non-spatial elements. Here there are two basic interpretations of that emergence. The first is that there are non-spatial elements at some scale below the Planck-length (where space simply does not exist) that, by being stitched together in the right way—through quantum entanglement for instance—produce space. And given the holographic principle, it seems that the space that emerges is 2-dimensional, as we've seen. But the other interpretation is that space does not emerge at all in external reality. On this interpretation, *external reality may be spaceless, and space may merely be the interface that we evolved* (to perceive fitness not truth). Both interpretations of the nature of space are consistent with the ITP. But what is interesting is that the second, perhaps more radical interpretation of the nature of space fits quite well with the ITP.

In physics, time is also inextricably bound up with space. And as we've seen, the most straightforward implication regarding the nature of time in special relativity (one of the bedrock frameworks in modern physics) is that the flow of time is illusory—that external reality is instead like a frozen "block" where everything just *is*, eternally, neither coming into existence nor fading out of it. The timeless theme is also something that pops out of the leading approaches to quantum gravity too.

There's another interesting hypothesis that can be formulated in light of duality and the ITP. Let's first assume that, as per the holographic principle, space is 2-dimensional, or, as per the main approaches to quantum gravity, space may not exist at all. Furthermore, let's assume for argument's sake that M-theory is the correct theory for completely uniting general relativity and quantum mechanics, and for uniting the four fundamental forces of nature (gravity, electromagnetism, and the strong and weak nuclear forces). Then it would be the case that the physics we observe could in principle be explained by different configurations of Calabi-Yau manifolds (the shapes of the very tiny extra dimensions of space). Let's further suppose that Calabi-Yau manifold *x* is the shape physicists would detect if, in principle, they could detect evidence for it empirically.

If the FBT Theorem and ITP are true, however, then had our perceptual interface evolved sufficiently differently, it would have been possible for physicists to have detected evidence for Calabi-Yau manifold *y instead* (assuming, again, that they could in principle detect evidence for it empirically). This is because, according to the ITP, our perceptions—spatial perception included—are not perceptions of external reality in and of itself; our perceptions are merely an interface for tracking fitness. So these counterfactual physicists could have evolved differently, and such that the perceptions they would have, if they could empirically detect evidence for the shape of the Calabi-Yau manifold of our universe, would be different: instead of finding evidence for shape *x*, they would find evidence for shape *y*. Either way, *neither* shape would be a true depiction of external reality. This is because, as per the ITP, perception has no direct access to external reality, and because external reality is actually 2-dimensional as per the holographic principle—or possibly even spaceless if one of the main approaches to quantum gravity is correct and if space doesn't actually emerge but is merely an interface that evolved. The view that Calabi-Yau manifolds may not actually exist in external reality, as a tangible part of the spatial fabric, is quite arguably an implication of the holographic principle: there we saw that everything in our universe actually lives on a 2-dimensional surface, and not within the volume of a 3-dimensional space. And we also saw that this 2-dimensional reality can nonetheless be given an equivalent description (via duality) in 3 dimensions. And Calabi-Yau manifolds are embedded—curled up—within 3 dimensions of space, and so they're merely part of that equivalent description.

According to this hypothesis, then, the dualities in string theory, such as the different shapes of Calabi-Yau manifolds and the strings and fluxes arrayed on them, are also not what external reality is like in and of itself; the dualities are merely ways of representing the connections between that external reality—whatever it is like, fundamentally—and the way that we evolved to perceive it. This would be analogous to glasses that come in a range of differently colored tints. Each of them represents the world in different colors, but they all interface with the same external world. So, dualities in string theory may indicate that, whatever external reality is like in and of itself, it is possible to represent it in more than one way. One key difference, however, is that whereas dualities in physics must in fact be precisely theoretically equivalent descriptions of what is observed (there can be no wiggle room), there may be a less demanding requirement in the case of the ITP. As we've seen, the ITP says that we didn't evolve to perceive reality, we evolved to perceive fitness. So there may be more ways for evolution to produce interfaces in perceivers than there are ways (for physicists, in particular) to discover theoretically equivalent descriptions of the way external reality is connected to what we observe.

On this view, much or most of science aims to give us *intersubjective truth*—that is, truths that similar perceivers can converge on, since they have similar perceptual faculties that evolved via natural selection (and because they can construct the same or similar conceptual maps [Churchland, 2012]). On the other hand, suppose we encountered an intelligent alien species, or synthetic intelligence of alien origin (robots with artificial intelligence, for example), that perceived the world rather differently than us—perhaps even radically differently. Nonetheless, we and the aliens should in principle be able to translate our differing perceptual experiences of reality, and our differing scientific explanations of reality, into an abstract and invariant structural explanation of some kind. This would be a kind of explanation that we could both recognize as shared—one that could be translated into and rendered meaningful within each of our experiences of reality. Despite our different experiences and understandings of reality, then, this abstract and invariant explanation of reality would also show how our differing perceptions and scientific explanations are nonetheless equivalent in some sense.[9] "For example, we might bump into a rock lying on the ground, and the aliens would encounter that same rock—or whatever served as its counterpart in their perception. Even if the perceptual experience of bumping differed between us, both of us would recognize its presence through a shared, abstract framework—an explanation that translates our distinct experiences into a common language. Interestingly, the ITP might also explain why psychotropic drugs have consciousness-altering effects: the drugs might actually alter the evolved perceptual channels that we use to interface with external reality.

[9] Similar ideas to this approach can be found in group theory in mathematics and physics, and structuralism in the philosophy of physics. Seminal examples include Born (1953), Weyl (1950), and Auyang (1995).

Whether space is real in external reality or not (or 2-dimensional or not), and whether time flows in external reality or not, external reality might be a kind of quantum computer.[10] One possibility is that we're inside a computer simulation, perhaps even inside a simulation nested within any number of simulations, like a Matryoshka doll (Bostrom, 2003; Chalmers, 2022a). In the scenario in which something like Hoffman's theory and the simulation hypothesis are both true, it may be that the reality in which the simulation is housed is nothing at all like how we perceive that simulation. The same could be true if we're in one simulation nested inside another simulation, or nested inside many other simulations: the basement-level reality which houses all of the simulations might, for all we know, be very unlike how we perceive the particular simulation that we're in. But the ITP is also consistent with the radical and interesting view defended by Tegmark (2014): namely that reality is fundamentally a vast mathematical structure, and that objects such as tables, chairs, protons, and quarks are not ultimately tangible, but rather derivative manifestations of that underlying mathematical reality. According to this Platonic view, every possible mathematical structure is realized. In our case, our observable universe is a timeless mathematical structure that ultimately can be described at the most fundamental level by a theory of quantum gravity—but our universe is merely a tiny fragment of the full expanse of mathematical reality, which is immensely large. A view like Tegmark's would also explain Eugene Wigner's (1990) query about the "unreasonable effectiveness of mathematics" in science. The answer to the question of why mathematics is "unreasonably effective" would be: because reality *is* fundamentally just mathematics.

Although I'm agnostic on Hoffman's view, I do not find two parts of it persuasive. Namely, I do not find his view that external reality is fundamentally made up of conscious agents compelling, nor do I think his favored version of quantum mechanics is compelling. Firstly, I think that Everett's many-worlds is the most straightforward understanding of quantum mechanics (Carroll, 2020). And if the ITP is true, then I find it much more likely that consciousness is fundamentally a pattern of information that can, and a pattern of information that appears to us as a certain sort of brain activity. In this way, and unlike Hoffman, I think that approaches to consciousness such as the global-workspace theory (Dehaene, 2014; Baars et al., 2013) and attention-schema theory (Graziano, 2019; Graziano et al., 2020) are on the right track to explaining consciousness. Accordingly, the bet I would make is that "qualia"—the subjective "what it's like" of conscious experience—can be explained in terms of cognitive neuroscience, one way or another (Carruthers, 2019; Graziano, 2020; Dennett, 2019). Of course, if the ITP is true, explaining consciousness in terms of cognitive neuroscience would only be our way of understanding it in terms of our evolved interface; it would still remain an open question precisely what consciousness corresponds to in external reality. But even in that case, it need not be, as Hoffman hypothesizes, that reality is literally made up of conscious agents. In which case, and as we've already discussed, one of the possibilities would be a quantum-mechanical description, as per the holographic principle, or the spaceless view of quantum gravity. Interestingly, both of these possibilities can be reduced to Tegmark's view that all of reality is a mathematical structure.

[10] The idea of reality as a quantum computer is an implication of the holographic principle. Lloyd (2006) also argues for the view that the universe is essentially a quantum computer.

## Conclusion

This paper described Donald Hoffman's (2019) theory that we don't perceive reality: spacetime, objects, colors, sounds, tastes, and so forth, are all merely an interface that we evolved to track evolutionary fitness rather than to perceive truths about external reality. My main goals have been to buttress and extend his argument by exploring ideas in physics. The journey went through the exotic terrain of black holes, the holographic principle, string theory, duality, quantum gravity, and special relativity. I am skeptical of Hoffman's theory, but it is very interesting and merits further research.